\def\aa{{A\&A}}

\def\aj{{AJ}}
\def\annrev{{ARA\&A}}
\def\apj{{ApJ}}
\def\apjs{{ApJS}}

\def\mnras{{MNRAS}}

\def\pasp{{PASP}}
\newcommand \kms          {\rm{\hbox{km s$^{-1}$}}}

\newcommand \Lya          {\hbox{Ly$\alpha$}}

\newcommand \Msun          {\hbox{M$_{\odot}$}}

\newcommand \zaz          {{$z_a\kern -1.5pt \approx\kern -1.5pt z_e$}}
\newcommand \zllz         {{$z_a\kern -3pt \ll\kern -3pt z_e$}}
\newcommand \Zsun          {\hbox{Z$_{\odot}$}}

\documentclass[cup5b]{caps}
\usepackage{graphicx}
\usepackage{amssymb}
\usepackage{ociwsymp4}   %Use this for invited papers.

\HeadText{F. Hamann, M. Dietrich, B. Sabra, C. Warner}
\begin{document}

\pagenumbering{arabic}

%Author names should be in captital letters
%\author[]{F. HAMANN, M. DIETRICH, B. SABRA\\University of Florida}

%Example for multiple authors:
%
\author[]{F. HAMANN$^{1}$, M. DIETRICH$^{2}$, B. SABRA$^{3}$,
and C. WARNER$^{1}$
\\
(1) University of Florida, Gainesville, USA\\
(2) University of Florida, Gainesville, USA, and Georgia State University,
Atlanta, USA\\
(3) University of Florida, Gainesville, USA, and American University of
Beirut, Lebanon}

\chapter{Quasar Elemental Abundances \\ and
Host Galaxy Evolution}

\begin{abstract}
High redshift quasars mark the locations where massive galaxies are
rapidly being assembled and forming stars. There is growing evidence
that quasar environments are metal-rich out to redshifts of at least
five. The gas-phase metallicities are typically solar to several
times solar, based on independent analyses of quasar broad emission
lines and intrinsic narrow absorption lines. These results suggest
that massive galaxies (e.g., galactic nuclei) experience substantial
star formation before the central quasar becomes observable. The
extent and epoch of this star formation (nominally at redshifts
$z>2$, but sometimes at $z>5$) are consistent with observations of
old metal-rich stars in present-day galactic nuclei/spheroids, and
with standard models of galactic chemical evolution. There is further
tantalizing (but very tentative) evidence, based on FeII/MgII broad
emission line ratios, that the star formation usually begins
$\geq$0.3 Gyr before the onset of visible quasar activity. For the
highest redshift quasars, at $z\approx 4.5$ to $\sim$6, this result
suggests that the first major star formation began at redshifts
$\geq$6 to $\geq$10, respectively.
\end{abstract}

\section{Introduction}

Quasars are no longer perceived merely as exotic high-redshift
monsters. Rather, they are commonplace, in the sense that every
massive galaxy today was at one time an active quasar host. Quasars
are thus valuable probes of galaxy evolution. They light up the
surrounding gas (in young galactic nuclei), and they provide a bright
emission source for absorption line studies, during a poorly
understood, early phase of galaxy assembly. This review describes
measurements of the gas-phase elemental abundances near quasars, and
the implications for massive galaxy evolution. Please see Hamann
\& Ferland (1999, hereafter HF99) for a more comprehensive review of
this topic.

\subsection{Quasars, Galaxies, \& High-Redshift Star Formation}

Recent studies show that quasars, or more generally Active Galactic
Nuclei (AGNs), are natural byproducts of galaxy formation. The
super-massive black holes (SMBHs) that power AGNs are not only common
in the centers of galaxies, but the SMBH masses correlate directly
with the mass of the surrounding galactic spheroid (Merrit \&
Ferrarese 2001, Gebhart et al. 2000, Tremaine et al. 2002). Whatever
processes lead to the formation of galactic spheroids, e.g.,
elliptical galaxies and the bulges of grand spirals, must also
(somehow) create a central SMBH with commensurate mass. Most of the
SMBHs in galaxies today are ``dormant," or nearly so, because the
mass accretion has declined or ceased. But all of the galaxies
hosting SMBHs today must have been at one time ``active,'' with the
bright AGN phase corresponding to the final growth phase of the SMBH.

The bright AGN phase is expected to be brief. In the standard
paradigm of AGN energy production, the mass accretion rate,
$\dot{M}$, needed to maintain a given luminosity, $L$, is $L =
\eta \dot{M}c^2$, where $\eta$ is an efficiency factor believed to
be of order 0.1. The luminosity can also be expressed as a fraction
of the Eddington limit, $L = \gamma\, L_E \approx (1.5 \times
10^{46})\, \gamma M_8$ ergs s$^{-1}$, where $L_E$ is the Eddington
luminosity, $M_8$ is the SMBH mass relative to $10^8$ \Msun , and
$\gamma$ is a constant typically between 0.1 and 1 for luminous
quasars (Vestergaard 2003, Warner et al. 2003). If the bright AGN
phase corresponds to the final accretion stage where the SMBH roughly
doubles its mass while accreting at $\sim$50\% of the Eddington rate
($\gamma \approx 0.5$), then the lifetime of this phase should be
$\sim$$6\times 10^7$ yr (based on the expressions above). This
nominal lifetime fits well with the population demographics. In
particular, the observed space density of present-day SMBHs, with
masses from $<$$10^{6}$ \Msun\ to $\sim$10$^{10}$ \Msun , can account
for the high-redshift quasar population if every one of these SMBHs
shined previously as an AGN for a few $\times 10^7$ yr (Ferrarese
2002).

Another important aspect of the AGN--galaxy relationship is the close
link between AGNs and vigorous star formation. For example,
high-redshift quasars are often strong sources of sub-mm dust
emission, which is attributed to powerful starbursts in the
surrounding galaxies (Omont et al. 2001, Cox et al. 2002). Several
authors have noted that the growth in the quasar number density with
increasing redshift matches well the increasing cosmic star formation
rate (e.g., Franceschini et al. 1999). Quasars flourished at a time
(corresponding to $z\sim 2$ to 3) when most massive galactic
spheroids (e.g., elliptical galaxies and the bulges of grand spirals)
were frantically forming most of their stars (Renzini 1998, Jimenez
et al. 1999, Dunne et al. 2003). It seems likely that the same
processes that dump matter into galactic nuclei to form an SMBH also
induce substantial star formation.

All of this evidence indicates that high-redshift quasars are bright
beacons marking the locations where massive galactic nuclei are being
assembled --- vigorously making stars and building a central SMBH.
However, we may still have something like the old ``chicken versus
egg'' problem. Which came first, the galaxy or the quasar? The bulk
of the stars or the central SMBH? How mature are the host galaxies
when the central AGN finally becomes visible?

\section{Quasar Abundance Studies}

Quasar abundance studies seek to examine quantitatively the chicken
versus egg problem. How chemically enriched are quasar environments
and, by inference, how much star formation preceded the observed
quasars? Does the degree of enrichment depend on the type of AGN or
on properties (mass?) of the surrounding host galaxy? When does the
star formation begin in quasar environments? Are the host
environments of high redshift quasars less evolved and therefore less
metal rich? The answers to these questions will provide unique
constraints on high-redshift star formation and early galaxy
evolution.

Most of the effort in this field has been to measure the overall
gas-phase metallicities near quasars. But there is still more we can
learn from the relative metal abundances. The ratio of iron to
$\alpha$ elements is of particular interest because the $\alpha$
elements, such as O, Ne, and Mg, derive exclusively from massive-star
supernovae (Types II, Ib and Ic), while Fe has a dominant
contribution from intermediate-mass stars via Type Ia supernovae
(Yoshii et al. 1996). The SN~Ia contribution of Fe is delayed
relative to the SN~II+Ib+Ic products because of the longer lifetimes
of SN Ia precursors (integrated over a stellar initial mass
function). The amount of the delay is often assumed to be $\sim$1
Gyr, but Matteucci \& Recchi (2001) showed that this number depends
on environmental factors such as the star formation rate and initial
mass function. They argue that $\sim$0.3 Gyr is more appropriate for
young elliptical galaxies/spheroids (with high star formation rates).
In any case, the delay can be used as an approximate ``clock'' to
constrain the formation times of stellar populations. For example,
observations of large gas-phase Fe/$\alpha$ ratios near metal-rich
quasars would indicate that their surrounding stellar populations are
already $\geq$0.3 Gyr old.

\subsection{Spectral Diagnostics}

The first step is to understand the abundance diagnostics in quasar
spectra. The signature features are the broad emission lines (BELs),
with profile widths greater than 1000 -- 1500 \kms . Reverberation
mapping studies show that quasar BELs form in photoionized gas within
$\sim$1 pc of the central continuum source (Kaspi et al. 2000). Some
of the first quasar studies noted that BELs identify the same variety
of elements (hydrogen, carbon, nitrogen, oxygen, etc.) seen in much
less exotic stellar and galactic environments (Burbidge
\& Burbidge 1967). In particular, there are no obvious abundance
``anomalies.'' The first quantitative estimates of BEL region
abundances (see Davidson \& Netzer 1979 for an excellent review)
confirmed the lack of abundance anomalies compared to solar element
ratios, and showed further that the metallicities are probably within
a factor of $\sim$10 of solar.

More recent abundance studies include absorption lines among the
diagnostics. Quasar absorption lines are classified generically
according to their full widths at half minimum (FWHMs). Roughly 10\%
to 15\% of optically selected quasars have classic broad absorption
lines (BALs), with FWHM $>$ 2000 to 3000 \kms . At the other extreme
are the narrow absorption lines (NALs), which have typically FWHM $<$
a few hundred \kms . A useful working definition of the NALs is that
the FWHM is not large enough to blend together important UV doublets,
such as C IV $\lambda\lambda$1548,1551, whose velocity separation is
500 \kms . Intermediate between the NALs and BALs are the so-called
mini-BALs. The BALs and mini-BALs appear exclusively blueshifted with
respect to the emission lines, with velocity shifts from near 0 \kms\
to $>$30,000 \kms\ in some cases. Their broad and smooth profiles at
blueshifted velocities clearly identify outflows from the central
quasar energy source. NALs are also frequently blueshifted, but they
can appear at redshifted velocities up to $\sim$2000 \kms . NALs
within $\pm$5000 \kms\ of the emission line redshift are also called
``associated'' absorption lines (AALs). In general, NALs can form in
a wide range of environments, from high speed outflows like the BALs
to cosmologically intervening clouds or galaxies having no relation
to the quasar. One challenge in using NALs for abundance work is to
understand the location of the absorbing gas.

The following sections outline the procedures and results related to
each of these diagnostics. Please see HF99 for a more complete
discussion.

\section{Intrinsic Narrow Absorption Lines (NALs)}

Figure 1 shows some typical examples of associated NALs in the
spectrum of a bright quasar, PG~0935+417 with emission-line redshift
$z_{em} =1.966$. The AALs in this case have FWHMs from $\sim$12 to
$\sim$133 \kms , and they are blueshifted by $\sim$1400 \kms\ to
$\sim$3000 \kms\ with respect to $z_{em}$.

  \begin{figure}
    \centering
    \includegraphics[width=11cm,angle=0]{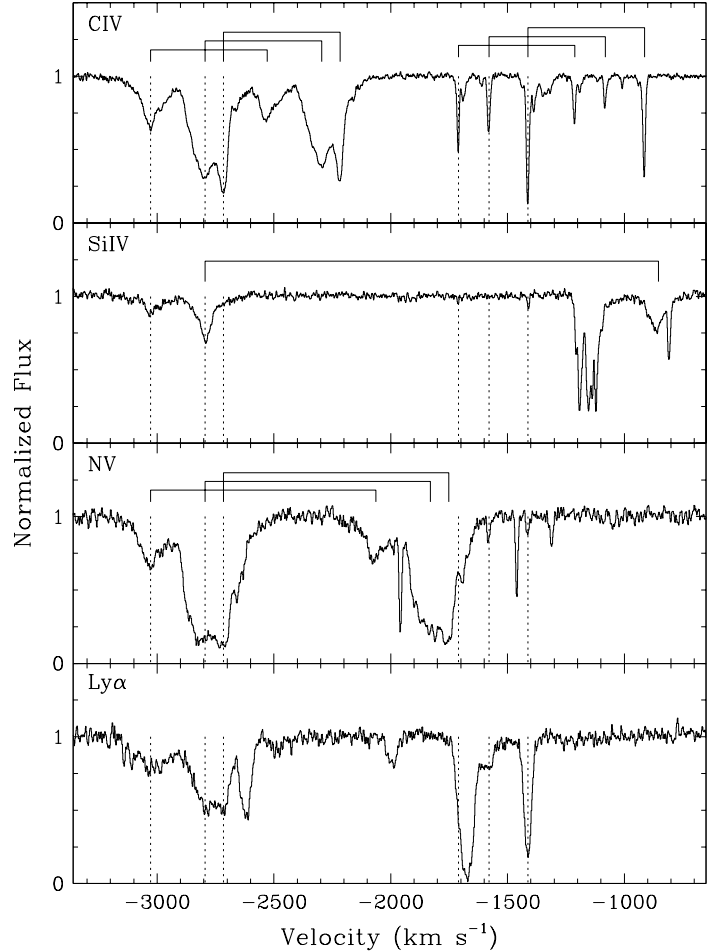}
    \caption{Normalized Keck--HIRES spectra of ``associated'' NALs in
the quasar PG~0935+417. Strong doublets are labelled above with open
brackets: C IV $\lambda\lambda$1548,1551, N V
$\lambda\lambda$1238,1242, and Si IV $\lambda\lambda$1393,1403. The
velocity scale is relative to $z_{em}$ = 1.966 for the short
wavelength component of each NAL doublet. The dotted vertical lines
mark the velocities of the six strongest C IV systems. }
    \label{fig1}
  \end{figure}

There are three critical steps involved in using NALs as abundance
diagnostics. First, we must distinguish the {\it intrinsic} NALs from
those that form in unrelated, cosmologically intervening gas. We
define ``intrinsic" NALs loosely as forming in gas that is (or was)
part of the overall AGN/host galaxy environment. Statistical studies
suggest that the AALs are often intrinsic to the quasar/host galaxies
(Richards et al. 1999, Foltz et al. 1986). Other observational
criteria must be applied to determine which specific NALs are
intrinsic, such as, i) time-variable line strengths, ii) NAL profiles
that are broad and smooth compared to the thermal line width, iii)
excited-state absorption lines that require high gas densities, and
iv) NAL multiplet ratios that reveal partial line-of-sight coverage
of the background light source (see Hamann et al. 1997a, Barlow \&
Sargent 1997). Characteristics like variability, broad-ish line
profiles, and partial covering often appear together, clearly
indicating an intrinsic origin. Some NALs show none of these
characteristics and the location of their absorbing gas remains
ambiguous.

Second, we must derive column densities in a way that can account for
an absorber that might be inhomogeneous and/or partially covering the
background light source(s). In the most general case, the observed
line ``intensities'' represent an average of the intensities
transmitted over the projected area of an extended emission
source(s). If the absorber happens to be homogeneous (has a constant
line optical depth) across the face of a uniformly bright emission
source, then the observed intensity is simply,
\begin{equation}
{{I(\lambda )}\over{I_o(\lambda )}} \ = \ 1\, -\, C_f(\lambda )\, +
\, C_f(\lambda )\, e^{-\tau_{\lambda}}
\end{equation}
where $I_o(\lambda )$ is the intensity of the emission source at
wavelength $\lambda$, $\tau_{\lambda}$ is the line optical depth,
$C_f(\lambda )$ is the coverage fraction of the absorbing material
($0\leq C_f(\lambda )\leq 1$), and we ignore small contributions of
line emission from the absorbing gas. If we measure two absorption
lines having a known optical depth ratio, e.g., in a doublet, then we
can use the measured strengths and ratios of those lines (providing
two equations) to solve uniquely for both $\tau_{\lambda}$ and
$C_f(\lambda )$ at each wavelength (Petitjean et al. 1994, Hamann et
al. 1997, Barlow \& Sargent 1997). For resonance lines, the ionic
column densities then follow simply from the optical depths, $N_{ion}
\propto\ \int\tau_{\lambda}\, d\lambda$. The key is to obtain
high-resolution spectra (so there are no unresolved line components)
of NAL multiplets. Fortunately, there are many possibilities, such as
the doublets shown in Figure 1.1, the H~I Lyman series lines, and
others.

If the absorbing medium is inhomogeneous, then more lines are needed
to define the two-dimensional optical depth/column density
distribution. However, a simple doublet analysis still provides a
useful estimate of the optical depth spatial distribution (at each
wavelength), subject to an assumed functional form for that
distribution. We recently completed an extensive theoretical study of
the effects of inhomogeneous absorption on observed absorption lines
(Sabra \& Hamann 2003), which builds upon the pioneering work of
deKool et al. (2001). We find that, for a wide range of inhomogeneous
optical depth distributions, the spatially averaged value of the
optical depth is very similar to the single value one would derive
(from the same data) assuming homogeneous partial coverage (Eqn.
1.1). Therefore, the most general situation effectively reduces to
the simple homogeneous case for the purposes of an abundance
analysis\footnote{There are second order effects that we will not go
into here, but it boils down to making a reasonable assumption about
the functional form of the column density/optical depth spatial
distribution.}.

The final step is to convert ratios of ionic column densities into
abundance ratios using ionization corrections. The correction factors
can be large because NALs are often highly ionized and we must
compare a high ion, such as C IV, to H I Ly$\alpha$ to estimate the
C/H abundance (metallicity). There can also be a range of ionizations
present in the absorber but not enough measured lines to fully
characterize this range. Nonetheless, Hamann (1997) showed that we
can always derive conservatively low estimates of the metal/hydrogen
abundance ratios by assuming each metal line forms under conditions
that most favor that ion. In particular, we can use the maximum
possible value of the metal ion fractions, such as C IV/C, as
indicated by photoionization calculations. Similarly, there are
minimum ionization corrections for each metal ion that provide firm
lower limits on the metal/hydrogen abundance (see also Bergeron \&
Stasinska 1986). These conservative estimates and firm lower limits
provide useful constraints on the metallicity even if we we have no
knowledge of the degree or range of ionizations in the gas (see also
HF99).

\subsection{Results}

High resolution spectra suitable for NAL abundance studies did not
become available until the 1990s. The earliest results (Wampler et
al. 1993, Petitjean et al. 1994, among others) were confirmed by
later studies (Hamann 1997, Hamann et al. 1997a, Tripp et al. 1996,
Savage et al. 1998, Petitjean \& Srianand 1999, Papovich et al.
2000). Quasar AALs often have metallicities $Z\geq$ \Zsun . Several
of these studies also report N/C above solar, although in most cases
the data are not adequate for this measurement. So far there are only
two published reports (to our knowledge) of AAL abundances above
redshift 4, and the results are slightly mixed. Wampler et al. (1996)
estimated $Z\sim 2$ \Zsun\ based on the tentative detection of O I
$\lambda$1303 in one AAL system at redshift 4.67, while Savaglio et
al. (1997) reported $0.1\leq Z\leq 1$ \Zsun\ for an AAL complex at
redshift 4.1. However, in both of these cases, and many others, the
location of the absorber is not known. All of the confirmed intrinsic
NALs (based on the criteria mentioned \S1.3) have metallicities
$Z\geq$ \Zsun\ (see refs. above). Petitjean et al. (1994) used a
small sample of redshift $\sim$2 quasars to show that there is a
dramatic decline in absorber metallicity as one moves away from the
quasar emission redshift, from $Z\geq$ \Zsun\ in the AALs to $Z <
0.1$ \Zsun\ at velocity shifts $\geq$10,000 \kms\ (see their Figure
14, also Carswell et al. 2002). This result meets our expectation
that AALs are often intrinsic to quasars, while most NALs at lower
redshifts are unrelated (Rauch 1998).

There is a strong need now to improve on some of the early
measurements and expand the database. We and collaborators have
obtained high-resolution (7--9 \kms ) spectra of 15 AAL quasars at
$z_{em}\approx 2$ -- 3 using the HIRES echelle spectrograph on the 10
m Keck I telescope. As a brief example consider the AALs of quasar
PG~0935+417 shown in Figure 1.1. We find no evidence of line
variability (above $\sim$10\%) based on several observations that
span $\sim$2 years in the quasar rest frame. Nonetheless, the three
systems of C IV doublets at 3000 to 2700 \kms\ are clearly intrinsic
because of the large non-thermal line widths and doublet ratios that
imply partial line-of-sight coverage of the background light source
(see also Hamann et al. 1997b). The location of the three narrower
absorption systems at 1700 to 1400 \kms\ remains unclear. However, a
preliminary analysis of all of these lines (based on simple
equivalent width measurements and assuming constant coverage
fractions across the profiles) suggests that the metallicities in all
cases are $\sim$1 to $\sim$3 \Zsun . A more detailed analysis of this
data set is underway, but solar or higher metallicities appear to be
common in the AALs, in agreement with earlier studies.

\section{Broad Absorption Lines \& Mini-BALs}

Figure 1.2 shows the rest-frame UV spectrum of a fairly typical BAL
quasar, PG 1254+047. The BALs in this case imply outflow velocities
from roughly 15,000 to 27,000 km s$^{-1}$. The main obstacle to using
BALs for abundance work is that the broad profiles blend together all
the important doublets. Therefore, unlike the NALs, we cannot examine
the doublet ratios for evidence of partial line-of-sight coverage. If
we simply assume that the coverage fraction is unity ($C_f = 1$ in
Eqn. 1.1), then measurements of the column densities imply bizarre
abundance ratios, such as, [Si/C] $>$ 0.5 and [C/H] $\approx 1$ to 2
(where [$x/y$] = $\log(x/y) - \log(x/y)_{\odot}$). The surprising
detections of P~V $\lambda\lambda$1118,1128 BALs in at least half of
the well-measured sources, including PG~1254+047 (Fig. 1.2), imply
[P/C] $\geq$ 1.6 (Junkkarinen et al. in prep., Hamann 1998, Hamann et
al. 2002a, and refs. therein).

 \begin{figure}
    \centering
    \includegraphics[width=7cm,angle=0]{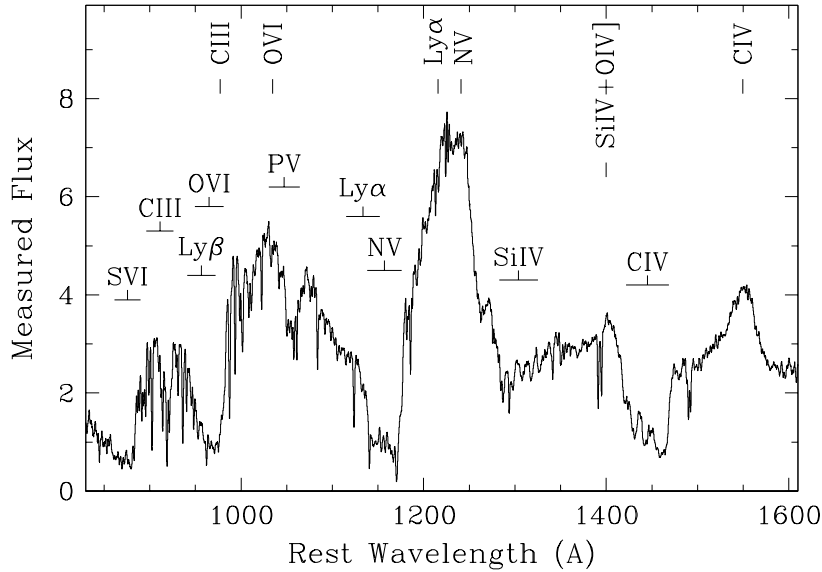}
    \caption{Hubble Space Telescope spectrum of PG~1254+047 ($z_{em}
= 1.01$) showing its strong BAL troughs. The BALs are labelled just
above the spectrum, while the prominent broad emission lines are
marked across the top. The measured flux has units 10$^{-15}$
ergs/s/cm$^2$/\AA . From Hamann (1998).}
    \label{fig3}
 \end{figure}

The most likely explanation for these strange abundances is that they
are incorrect. BALs are often much more optically thick than they
appear because the absorber partially covers the background light
source(s). Direct evidence for partial coverage has come from
comparisons of widely spaced line pairs (similar to the doublet
analysis but not involving doublets) in one well-measured BAL quasar
(Arav et al. 2001). Doublet ratios in borderline NALs/mini-BALs,
sometimes embedded within BAL profiles and believed to form in the
same general outflow, also frequently indicate partial coverage
(Hamann et al. 1997b, Barlow, Hamann \& Sargent 1997, Telfer et al.
1998, Srianand \& Petitjean 2000). Less direct evidence comes from
spectropolarimetry of scattered light in BAL troughs (Schmidt
\& Hines 1999), and from flat-bottomed BAL profiles that ``look''
optically thick even though they do not reach zero intensity.

Another clue comes from the surprisingly strong P {\sc v} BALs.
Hamann (1998) noted that the P~V line should be nominally weak. Its
ionization is very similar to C~IV, but in the Sun phosphorus is
$\sim$1000 times less abundant than carbon. Therefore, if the P/C
abundance is even close to solar in BAL regions, the P~V line should
not be present {\it unless} C~IV and the other strong BALs of
abundant elements are much more optically thick than they appear. In
other words, there is partial coverage and unabsorbed light fills in
the bottoms of BAL troughs. If we turn that argument around and
assume solar P/C in PG~1254+047 (Fig. 1.2), we find that the true
optical depth in its C IV line is $\geq$25 (Hamann 1998).

This interpretation of the P~V may be disputed by Arav et al. (2001),
who measured P~V and many other BALs in one quasar spectrum with
extraordinarily wide wavelength coverage. They estimate a metallicity
near solar but [P/C] $\geq$ 1. If that P/C result is correct, it
would probably require a unique enrichment history (e.g., involving
novae, Shields 1996). However, the uncertainties and challenges are
substantial. We prefer to exclude BALs from quasar abundance studies.

\section{Broad Emission Lines}

Figure 1.3 shows part of the rest-frame UV spectrum of a quasar at
redshift $z_{em}=4.16$. The emission lines shown in this plot, plus
C~III] $\lambda$1909, have formed the basis for many BEL metallicity
studies. The main advantage of the BELs is that they can be measured
in any quasar using moderate resolution spectra. There is also no
question that the lines form close to the quasar, nominally within
$\sim$1 pc of the central engine (\S1.2.1).

One issue affecting the BEL analysis is that the emitting regions
span simultaneously a wide range of densities and ionizations, with
the higher ionizations occurring preferentially nearer the central
continuum source (e.g., Ferland et al. 1992, Peterson 1993).
Consequently, different lines can form in spatially distinct regions.
Without a detailed model of the BEL environment, it is important for
abundance studies to compare lines that form as much as possible in
the same gas with similar excitation and radiative transfer
dependencies. However, the range of physical conditions present in
BEL regions provides a simplification: observed BEL spectra are
flux-weighted averages over a diverse ensemble of ``clouds.'' The
tremendous advantage of this natural averaging is that we do not need
to derive, or make specific assumptions about, the physical
conditions in the different line emitting regions (Hamann et al.
2002). The formalism developed to simulate this situation (Baldwin et
al. 1995) has been dubbed the Locally Optimally-emitting Cloud (LOC)
model, because each line forms naturally where the conditions most
favor its emission.

 \begin{figure}
    \centering
    \includegraphics[width=11.5cm,angle=0]{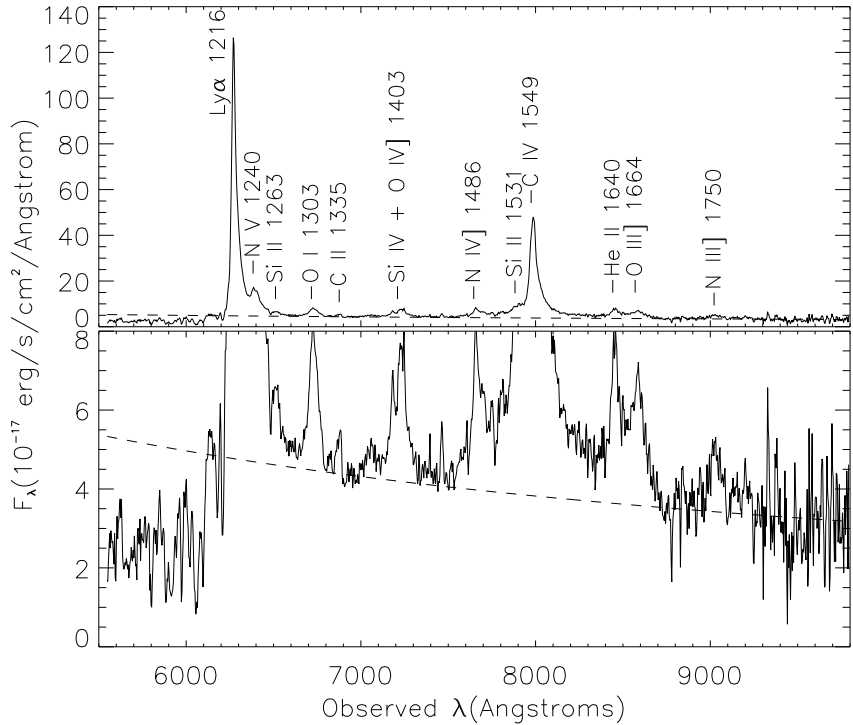}
    \caption{Broad emission lines in the rest frame UV spectrum of the
redshift 4.16 quasar, BR 2248$-$1242. The lower panel is an expanded
version of the upper plot. From Warner et al. (2002).}
    \label{fig4}
 \end{figure}

Another important consideration is that the combined emission in the
metal lines is not sensitive to the overall metallicity. In
particular, the strengths of prominent metal lines, such as C IV,
relative to the hydrogen lines, such as Ly$\alpha$, are not sensitive
to the metal-to-hydrogen abundance ratio (HF99). The main reason is
that these BELs are the dominant coolants in the photoionized plasma
where they form. Radiative equilibrium requires that the total energy
emitted from any region in the plasma equals the total energy
absorbed. Therefore, changing the metallicity by factors of several
cannot produce a commensurate change in the overall line strengths
without violating energy conservation.

Nonetheless, BELs are sensitive to the metal abundances in several
ways (see Ferland et al. 1996, HF99, Hamann et al. 2002). First,
departures from solar metallicity by orders of magnitude {\it will}
change the total metal/H line emission ratio (as well as individual
line ratios such as C IV/\Lya ). For example, at very low
metallicities ($\leq$0.03 \Zsun ) the metal lines no longer control
the cooling and their emission strengths decline roughly commensurate
with the metal/H abundance. This fundamental sensitivity to the
extremes implies that typical BEL metallicities are conservatively
within a factor of $\sim$30 of solar. Second, weak metal lines that
are not important for the cooling do vary significantly with
metallicity (while still preserving the overall energy balance).
Abundance studies should, therefore, endeavor to include weaker
lines, such as O III] $\lambda$1663 and N III] $\lambda$1750 (Fig.
1.3). Finally, the {\it relative} strengths of different metal lines
can be sensitive to the metal/metal abundance ratios.

Taking advantage of this last point, Shields (1976) noted that in
galactic H II regions the N/O abundance scales roughly proportional
to O/H (metallicity) because of a strong ``secondary'' contribution
to the nitrogen enrichment (see also Pettini et al. 2002, Pilyugin et
al. 2003). The N/O $\propto$ O/H scaling dominates for metallicities
above a few tenths solar. The N/O abundance therefore provides an
indirect indicator of the overall metallicity. This technique has
become the norm for BEL metallicity studies.

It should be noted that departures from the simple N/O $\propto$ O/H
scaling can occur if there are metallicity-dependent stellar yields,
or if the enrichment is dominated by star formation in discrete
bursts (Kobulnicky \& Skillman 1996, Henry et al. 2000). The latter
situation would lead to time-dependent fluctuations in N/O because of
different delays in the stellar release of N and O. However, the
overall trend for increasing N/O with O/H remains, and the best
homogeneous data sets indicate that the scatter in the N/O $\propto$
O/H relation declines with increasing O/H (Pettini et al. 2002,
Pilyugin et al. 2003). This is the regime of quasars. Moreover, there
are no reports, to our knowledge, of high N/O ratios (solar or
higher) in metal poor (significantly sub-solar) interstellar
environments. Large N/O abundances are an indicator of high
metallicities in any scenario that involves well-mixed interstellar
gas. The situation with N/C can also be complicated because N and C
arise from different stellar mass ranges, leading again to
time-dependent effects (Henry et al. 2000, Chiappini et al. 2003).
Nonetheless, for the BEL analysis, it is generally assumed that to
first order N/C behaves like N/O.

\subsection{Results: Metallicity}

Shields (1976) analyzed several ratios of UV intercombination
(semi-forbidden) lines (see Fig. 1.3) and found that N/O and N/C are
nominally solar to $\sim$10 times solar, consistent with solar to
super-solar metallicities (see also Baldwin \& Netzer 1978, Davidson
\& Netzer 1979, Uomoto 1984). Hamann \& Ferland (1993) and
Ferland et al. (1996) later claimed that the metallicities are
typically several times solar, based on a new analysis of the
N~V/C~IV and N~V/He~II BEL ratios. More recently, Hamann et al.
(2002b) used extensive photoionization calculations to quantify
better the abundance sensitivities of all of these nitrogen line
ratios. In particular, they explored the influence of non-abundance
effects such as density, ionization, turbulence, and incident
continuum shape, in the context of LOC calculations. They favored N
III]/O III] and N V/(C IV + O VI $\lambda$1034) as the most robust
indicators of the relative N abundance. N~V/He~II is also very
useful, but it has a greater sensitivity to the ionizing continuum
shape because it compares a collisionally-excited line (N~V) to a
recombination line (He~II, see also Ferland et al. 1996). Holding
other parameters constant, the nitrogen line ratios scale almost
linearly with the N/O and N/C abundances and should, therefore, be
proportional to the metallicity if the enrichment follows N/O
$\propto$ O/H.

Dietrich \& Hamann (2003a) used the Hamann et al. (2002b)
calculations to estimate BEL metallicities in a sample of $\sim$700
quasars spanning redshifts $0\leq z_{em}\leq 5$. They find typically
several times \Zsun\ across the entire redshift range. Figure 1.4
shows results for a sub-sample of these quasars at $z_{em}\geq 3.5$,
where the average metallicity is $\sim$5 \Zsun\ (from Dietrich et al.
2003b). Notice that there is no evidence of a metallicity decrease at
the highest redshifts. For the particular $z_{em} = 4.16$ quasar
shown in Figure 1.3, where many lines are measurable owing to the
narrow profiles and large equivalent widths, Warner et al. (2002)
estimated $Z \approx 2$ \Zsun . Dietrich et al. (2003a and 2003b)
noted that the metallicities derived from the N~V lines ratios (most
notably N~V/He~II) are typically $\sim$30\% to a factor of $\sim$2
larger than estimates from the intercombination ratios (e.g.,
N~III]/O~III]). The reason for this discrepancy is not clear. It
could arise from systematic measurement errors in the weaker lines
(Warner et al. 2003), or, perhaps, from subtle excitation/radiative
transfer effects in the emitting regions. In any case, a factor of
$\leq$2 agreement among the different BEL diagnostics is quite good.
Our current strategy is to average the results from the various
nitrogen BEL ratios.

The absolute uncertainties are more difficult to quantify because
they depend on the theoretical techniques and assumptions. Factors of
a few can be expected, but the essential result seems secure. The
metallicities in quasar BEL regions are minimally near solar and
perhaps typically several times above solar. This result is confirmed
in general terms by the AAL data (\S1.3.1, see also Kuraszkiewicz \&
Green 2003).

 \begin{figure}
    \centering
    \includegraphics[width=10.5cm,angle=0]{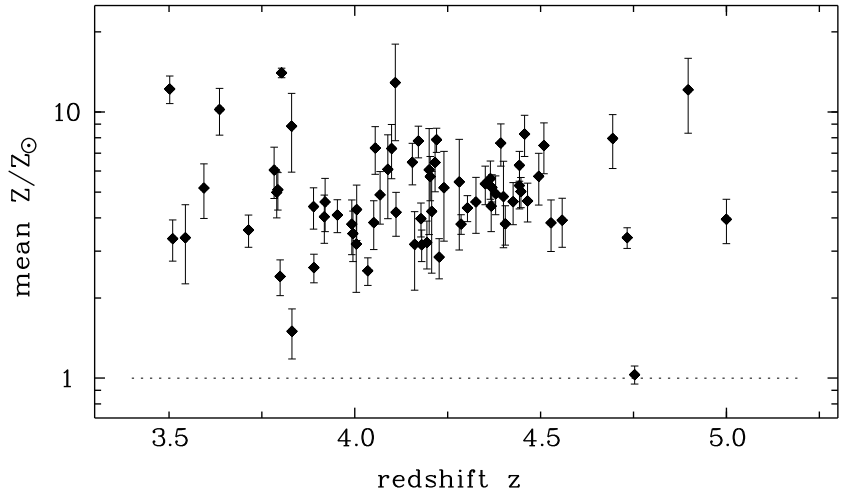}
    \caption{Average metallicities derived from BEL ratios
in a sample of high redshift quasars. The error bars represent only
the measurement uncertainties. From Dietrich et al. (2003b.)}
    \label{fig5}
 \end{figure}

Another interesting result from the BELs is that more luminous
quasars appear to be more metal rich (Hamann \& Ferland 1993, Osmer
et al. 1994, HF99, Dietrich et al. 2003a). This trend is somewhat
tentative because it is stronger in the N~V/C~IV and N~V/He~II ratios
than in the intercombination lines, and it has not yet been tested in
the AAL data. Nonetheless, it meets simple expectations. More
luminous quasars are powered by more massive SMBHs, which reside in
more massive galaxies (\S1.1). There is a well-known relationship
between mass and metallicity in normal galaxies (e.g., Trager et al.
2000), which is generally attributed to the effects of galactic
winds. Massive galaxies reach higher metallicities because they
retain their gas longer against the building thermal pressures from
stellar mass loss and supernova explosions. Therefore, the gas near
quasars in more massive galaxies could be more metal rich. Warner et
al. (2003) tried to test this idea by comparing BEL metallicities to
estimates of the SMBH masses (which serve as a surrogate for the host
galaxy mass) in the large quasar sample of Dietrich et al. (2003a).
The results are uncertain, but they do favor a trend between SMBH
mass and BEL metallicity, as expected. In fact, our best estimate for
the slope of this trend agrees well with the mass--metallicity slopes
observed in galaxies.

\subsection{Results: Fe/$\alpha$}

BELs are the only diagnostics used so far to estimate Fe/$\alpha$
abundances as an age discriminator(\S1.2). This work has relied on
Fe~II(UV)/Mg~II $\lambda$2798, where ``Fe~II(UV)'' represents a broad
blend of many Fe~II lines between roughly 2000 and 3000 \AA . This
blend is by far the most prominent Fe~II feature in quasar spectra,
and its UV wavelength makes it measurable at the highest redshifts in
the near infrared. The nearby Mg~II doublet serves the
$\alpha$-element representative with an ionization similar to Fe~II.

Measuring the flux in the broad Fe~II blend presents a unique
challenge (see Dietrich et al. 2003c for discussion). Spectra with
wide wavelength coverage are essential to properly define the
underlying continuum. It also helps to use a scaled template to fit
the Fe~II blend, based on either theoretical predictions (Wills et
al. 1985, Verner et al. 1999, Sigut \& Pradhan 2003) or observations
of a well-measured source (Vestergaard \& Wilkes 2001).

The main difficulty, however, is understanding the theoretical
relationship between the Fe~II/Mg~II emission ratio and the Fe/Mg
abundance. The Fe~II atom has hundreds of relevant energy levels and
thousands of lines that can blend together and/or contribute to
radiative pumping/fluorescence processes. Current interpretations of
observed Fe~II spectra are still based largely on the pioneering
calculations by Wills et al. (1985). Their work suggests that the
strong Fe~II fluxes from typical quasars require a relative iron
abundance (e.g., Fe/Mg) that is super-solar by a factor of roughly 3.
More modern calculations (Verner et al. 1999 and this proceedings,
Sigut \& Pradhan 2003, Baldwin et al. in prep.) use better atomic
data and can, for the first time, incorporate a many-level Fe~II atom
into fully self-consistent treatments of the energy budget. So far
these calculations have mostly just confirmed what Wills et al. knew
already, that the uncertainties are large. However, work is now
underway to understand these uncertainties. In particular, it will be
important to quantify the sensitivities of the Fe~II emission and the
Fe~II/Mg~II ratio to various poorly-constrained, non-abundance
parameters in BEL regions. It might turn out that the Fe~II(UV)/Mg~II
ratio is not the best Fe/$\alpha$ diagnostic (Verner et al. this
proceedings).

Putting these concerns aside for the moment, the observational
results are becoming increasingly interesting. The best data indicate
that the Fe~II/Mg~II emission ratio is the same on average at all
redshifts (Thompson et al. 1999, Iwamuro et al. 2002, Dietrich et al.
2002 and 2003c and this proceedings). This includes the most recent
observations of two quasars at redshifts $z_{em} \approx 6$
(Freudling et al. 2003). If we accept the tentative Wills et al.
(1985) result that Fe/Mg is nominally above solar in BELs, which they
deduced from spectra of quasars with $z_{em}\leq 0.5$, then the newer
data suggest that Fe/Mg is large at all redshifts. Therefore, SN Ia's
played a role in the enrichment and the star formation must have
begun at least 0.3 Gyr prior to the observed quasar epochs. For the
redshift $\sim$6 quasars, this line of argument implies that the
first major star formation occurred at redshift 10 -- 20 (Freudling
et al. 2003), consistent with the epoch of re-ionization deduced from
recent WMAP data (Bennett et al. 2003).

\section{Summary \& Implications}

Independent analysis of the BELs and intrinsic NALs indicates that
quasar environments are metal rich at all redshifts. Typical
metallicities range from roughly solar to several times solar. There
might also be a trend with luminosity in the sense that more luminous
quasars, which reside in more massive host galaxies, are more metal
rich. If the gas near quasars was enriched by a surrounding stellar
population, then the high metallicities indicate that those
populations are already largely in place by the time the quasars
``turned on'' or became observable. For example, simple
``closed-box'' chemical evolution with a normal stellar initial mass
function will produce metallicities above solar only after $\geq$60\%
of the original mass in gas is converted to stars. The quasar data
imply that this degree of enrichment and evolution is common in
massive galactic nuclei before redshift 2, and sometimes it occurs
before redshift 5.

Unfortunately, the quasar data do not tell us the size (mass) of the
stellar population responsible for the enrichment. However, the mass
of gas in the BEL region sets a lower limit. The most recent
estimates indicate that luminous quasars have BEL region masses that
are conservatively $\sim$10$^3$ to 10$^4$ \Msun\ (Baldwin et al.
2003). Normal galactic chemical evolution then suggests that the
stellar population needed to enrich this gas to $Z\geq$ Z$_{\odot}$
is ten times more massive, or $\sim$10$^4$ to 10$^5$ \Msun . This
mass may still be just the ``tip of the iceberg'' if one considers
that quasar BEL regions are part of a much more massive reservoir
that includes a $\geq$$10^9$ \Msun\ SMBH and its surrounding
accretion disk. BEL regions might, in fact, be constantly replenished
by the flow of material through the accretion disk. If that flow is
$\sim$10 \Msun\ yr$^{-1}$ of metal-rich gas (\S1.1), then clearly a
much more massive stellar population (perhaps with Galactic
bulge-like proportions) would be needed for the enrichment (see
Fria\c{c}a \& Terlevich 1998 for specific enrichment models). for
Observations of strong dust and sometimes CO emissions from
high-redshift quasars (currently up to $z_{em}\approx 4.7$) indicate
that there is often already $\geq$$10^{10}$ \Msun\ of metal-enriched
gas present (Omont et al. 2001, Cox et al. 2002). There must have
been massive amounts of star formation in these quasar host galaxies
before the observed quasar epoch. Efforts to date the stellar
populations around quasars using Fe~II/Mg~II BEL ratios suggest that
the star formation often begins in earnest $\geq$0.3 Gyr prior to the
appearance of a visible quasar. These results are tentative because
of theoretical uncertainties, but for the highest redshift quasars
yet studied ($z_{em}\approx 6$) they suggest that the first star
formation occurred at $z\sim 10$ -- 20.

In terms of the ``chicken versus egg" problem (\S1.1) the quasar
abundance data are clear; a substantial stellar population is already
in place by the time most quasars become visible. At low redshifts,
direct imaging studies of quasars and their lower-luminosity cousins,
the Seyfert 1 galaxies, show clearly that the host galaxies are
already present with substantial, even moderately old, stellar
populations on $>$ kpc scales (Nolan et al. 2001, Dunlop et al.
2003). At higher redshifts there is less direct imaging data, but at
least some quasars still have substantial hosts (Kukula et al. 2001).
It could be that the stars and the SMBH begin forming at the same
time. But {\it visible} quasar activity might be delayed with respect
to the surrounding star formation, even at the highest redshifts,
because of the time needed to assemble the SMBH and/or blow out the
dusty interstellar medium that obscures the youngest AGNs from our
view (see also Romano et al. 2002, Kawakatu \& Umemura 2003). This
delay could explain observations showing that the quasar number
density declines dramatically with increasing redshifts above $z\sim
3$, while the cosmic star formation rate appears to stay roughly
constant out to at least $z\sim 4$ (Ivison et al. 2002).

\noindent{\bf Acknowledgements.} \ We are grateful to our collaborators,
especially Jack Baldwin, Tom Barlow, Gary Ferland, Vesa Junkkarinen,
and Kirk Korista, who contributed insights and unpublished results to
this review. We also acknowledge NSF grant AST99-84040.

\end{document}